# AI & Racial Equity: Understanding Sentiment Analysis Artificial Intelligence, Data Security, and Systemic Theory in Criminal Justice Systems

**Alia Abbas**

**Systemic Justice Institute, Harvard Law School**

**Supervisor:** Professor Jon Hanson

**Word Count:** 6323

# Table of Contents

**Abstract**

Various forms of implications of artificial intelligence that either exacerbate or decrease racial systemic injustice have been explored in this applied research endeavor. Taking each thematic area of identifying, analyzing, and debating an systemic issue have been leveraged in investigating merits and drawbacks of using algorithms to automate human decision making in racially sensitive environments. It has been asserted through the analysis of historical systemic patterns, implicit biases, existing algorithmic risks, and legal implications that natural language processing based AI, such as risk assessment tools, have racially disparate outcomes. It is concluded that more litigative policies are needed to regulate and restrict how internal government institutions and corporations utilize algorithms, privacy and security risks, and auditing requirements in order to diverge from racially injustice outcomes and practices of the past.

## Introduction

In 2020, the groundbreaking cases of natural disasters and racial unrest shaped the world's daily routine, actions, and language. Oxford Languages & lexicographers, in their annual pick for the phrase that defines the past 12 months chose various words for the year of 2020, including "Black Lives Matter", "Coronavirus", "lockdown", but most importantly, "systemic racism". The latter phrase increased by 1,623% since 2019 in language use frequency (Oxner, 2020).

And yet, according to a nonpartisan national survey poll conducted by the University of Massachusetts Lowell Center for Public Opinion, the majority of Americans, especially white respondents, don't fully understand the disadvantages faced by Blacks to their advantages and the full extent of structural privileges at their favor, or simply, what systemic justice is. Only 33% of whites favor or strongly favor the creation of a citizen review board at the municipal level made of racially representative community members to review police actions, while Blacks stood at 42% in favor. 83% of Blacks polled that they feel discrimination in seeking a job, compared to 41% of whites who said both races had equal chance. In the education sector, 34% of Americans believed the field had equal opportunity for all children, while 81% of Black respondents said otherwise. Lastly, the poll found 57% of Americans are either undecided or opposed to systemic help, while 66% of Black respondents said it was needed. (Poll: American's Views of Systemic Racism Divided by Race, 2020).

The technical term, systemic, is often used only as a scientific phrase. Over the past 50 years, the term has moved from linguistic anthropology, to sociology, and now computational algorithms and natural language processing in machines and softwares that automate decisions. The importance for this paper is to highlight the sociological, psychological, anthropological, definitive & theoretical, algorithmic, and litigative definitive aspect of systemic theory for the 21st century. Since the 1980's, American social sciences has sophisticated and diversified its "color-blind, language-centered, racism denial tactics" and can make language processing algorithms used by machine learning vulnerable in picking up such bias as normal and extending it further into the digital space of the new higher intelligent machines, who will replicate the thinking (Thompson-Miller, 2017).

The purpose of this research paper is to investigate the topic of pre-existing racial inequality in the criminal justice system and the decision making and data processes executed by Artificial Intelligence to determine underlying bias and systemic injustice within outcomes of this management.

## Methodology

The aim of this study is to determine the identification and policy around systemic racism patterns that historically exist and are currently transitioning into automated practices with algorithms in the criminal justice system. Due to the focus on behavior and decision-making in this study, the methodology is separated into thematic areas as followed: implicit bias, historical language patterns, policy, and algorithm. Each thematic area is explored within the different sectors examined. Each thematic area is examined with a theoretical lens of the systemic theory in order to find patterns, as well as a focus group.

The dissertation uses a qualitative method of focus groups and historical research. Qualitative data examined includes sex (male or female), race, nationality, and criminal background. Conclusions are drawn from research on documentary evidence and pre-existing data sets. The main purpose of this research method is to gain an understanding of compounding discriminatory practices and how they evolve over time, as well as how such practices differentiate on the basis of color and race. Since the study is concerned with racial equity, the chosen methods of analysis are designed to evaluate the experiences of demographics based on color, particularly white and black American individuals.

In order to show the total lineage of how the focus groups have been affected over the course of history, the research method strongly uses critical theory of changing perspective to reach the ability of exposing the total lineage of a particular group based on color, their level in respect to the other race group, and conditions and experiences compared to the other race group. Therefore, secondary data that is examined had to fulfill the following criteria:

- Race affiliated, must include the white and black demographic
- Must be a study based in criminal justice and private prison industry, higher education, and healthcare based in the United States.
- Experienced or evidence is available of historical discimination in the pertaining study

Prior to drawing conclusions, each thematic area has a set of questions examined to further fulfill the above criteria. Each thematic area is as follows:

Historical language patterns:

-  Are there two or more instances in which this practice has been done and had discriminatory outcomes against the same race group consistently?
- Can the practice evolve over-time due partly to the necessity it serves to society and that the practice cannot be completely abandoned?

Policy:

- Can litigation benefit and make tools out of these patterns?
- Can legislative protections be implemented and required (even if the sector is prominently privatized or competitive)?

- Can the extent of protections be incorporated into comprehensive consumer privacy legislation?

Implicit Bias:

- Is there a natural aversion toward a particular group or person in the practice, or is neutral?
- Can randomized controlled trials create similar outcomes?

Algorithm:

- Can there or has there been consequences of algorithmic practices?
- Is there a superior group within the system that is readily accountable in protecting individuals from discriminatory impacts of automated systems?

**Chapter One: Linguistics to Litigative Language to Computational Language**

**What is systemic theory?**

The systemic theory was first studied in linguistic anthropology around the mid 20th century. Notably the theory was developed by linguist John Ropert Firth, whose approach to examining the current state of the English language was to use or identify semiotic anthropology. He defined systems because he was able to see signs or symbols repeating within sentences that were heavily based on human 'cultural theory of signs'.

"The name 'systemic' derives from the term 'system', in its technical sense as defined by Firth in 1957; system is the theoretical representation of paradigmatic relations, contrasted with 'structure' for syntagmatic relations" (Koerner, 1996, page 1996). Paradigmatic is a relation that holds between elements of the same category, such that an element can be substituted for one another and become a typical example of something, similar to that of a caste system or bureaucratic levels. Whereas syntagmatic applies to relations holding between elements that are combined with each other.

"Paradigmatic and syntagmatic relations together constitute the identity of an item within the system as a whole" (Koerner, 1996). Each person or institution within a system can be characterized by (1) where it's able to occur sequentially with others, its distribution, or (2) its substitution. Asking questions such as, 'Is it possible for this other very specific group to have experienced similar conditions if they replaced the current group with a system? Or, does the current group within the system have an inferior or superior priority over others groups?' allows one to label the outcome as either paradigmatic, as systemic, or syntagmatic, as structural.

In order to identify a group in a paradigmatic society or whole system, describing them "'consists in' locating it with respect to the rest" so it shows their 'total lineage of agnate forms' or the level in which their ascendants were placed within society (Koerner, 1996, page 273).

**Differentiations between similar theories**

Competing theories are often confused or mistakenly substituted for the systemic theory, such as the structural theory or systematic theory. To separate system and structure, "system is an internally organized whole where elements are so intimately connected that they operate as one in relation to external conditions and other systems" (Spirkin, 2018). There are patterns of certain types of behaviors, heuristics, and deeply rooted structures that seem more normal. All systems have defects, thus the rise of economic inequality and caste systems ultimately lead to racial inequality, gender inequalities, and others based on more individualistic identities.

To address the system, one must be able to address underlying patterns. As opposed to structure, which is a "composite whole, or an internally organized content" (Spirkin, 2018).

"Structure is not enough to make a system. A system consists of something more than structure: it is a structure with certain properties" (Sprikin, 2018). Functions, the way of life or purpose in serving, are what organize structures. For example, the function of sight organized the eye similar to how the function of managing increasing numbers of incarcerated individuals organized private prisons. To address a structure, one can determine and address the functions. This lacks perspective change, the critical postmodernism theory in order to see the reasoning or justness of a certain function from the differing views of other and unique groups experiencing, managing, influencing, or impacted by the function in one way or another.

Lastly, systematic and systemic can be differentiated as something being consistent and arranged as an organized or something that is done according to a system and an occurrence inside a system, respectively.

**Systemic Justice: Arrival into Sociology & Psychology**

The arrival of the linguistic theory in the broader spectrum of social sciences did not come until the establishment and development of sociology in the mid-19th century when racial theorizing about the nature of races and the causes of racial inequality studies began. Specifically, it wasn't until the late 20th century that a "host of new theories- the new racism- sought to explain the persistence of racial inequality in the post-civil rights era", that is including the systemic theory (Thompson-Miller, 2017).

"Conventional racial analysis remained underdeveloped and uncritical because of a comprehensive white graming of knowledge, and a mitigating and obscuring of institutional, structural, and systemic racism" (Thompson-Miller, 2017, page 164). "While efforts by marginalized social scientists in the 1960s and 1970s to catch up with the civil rights movement by articulating the large and robust definition of racism as instituionalized or systemic," race relations suffered from conceptually minimized views from American psychologists who strongly suggested that the cause of racial inequality and discrimination was merely and simply a fact of prejudice (Thompson-Miller, 2017, page 23).

However, the work of those sociologists have lived on and the term systemic justice was born quietly into the early 21st century in light of the transition of the racial injustices that continued widespread and quietly within institutions and practices in the U.S. Joe R. Feagin, one of the most prominent racial social theorists and sociologist who primed the systemic theory into the field, defines systemic racism as what "encompasses a broad range of white-racist dimensions: the racist ideology, attitudes, emotions, habits, actions, and institutions of whites in this society...and is far more than a matter of racial prejudice and individual bigotry. It is a material, social, and ideological reality that is well-embedded in major US institutions" (Thompson-Miller, 2017, page 165).

Systemic racism is a corrective term coined by sociologists, well-informed civil rights leaders, & those who have experienced the phenomenon of 'awakening'. The term provides a more theoretical aspect that "fully grasps the complexities and contradictions" of the former incorrect, reductionist racial theories that existed during the civil rights era (Thomson-Miller, 2017). The old theories suggested that as the older prejudiced majority of whites died, they would be replaced with more racially tolerant liberal whites that would enable individual bigotry and racial inequality to die off as well.

**Understanding and 'realizing' systemic injustice: "The Awakening"**

"Systemic theory retains…'realization' is the relation between the 'strata' and the phases of representation within one stratum" (Koerner, 1996). The realization is when one is able to recognize relations of the 'strata', a level within a system, to one's own multistratal symbols (who or what represents each level) within a society. The relation between "language and other sociocultural phenomena is then modeled that of realization" (Koerner, 1996, page 273). Despite the theory being derived from a linguistics concept, it is understood that there are sociocultural phenomena occurring in order to identify or experience an 'awakening' or 'realization'. Linguists experienced the "awakening" as that paradigmatic, or systems exist. Specifically, "syntagmatic organization is interested in the 'realization' of paradigmatic features" (Koerner, 1996, page 273).

The realization sociologists and civil rights academics experience is the "awakening to the reality that the criminal-justice system [among other systems] is not just another institution infected with racial bias, but the primary engine of racial inequality and stratification in the U.S. today" (Alexander, 2010, xlix).

**Systemic Racism: Prolonged existence**

In the 21st century, especially, racial profiling and racism is a organized, lucrative, & secretive system of "race"- justified oppression. "Systemic white racism" is the "central and enduring social structure around which the United States and other modern societies are organized and evolved" that maintains the privilege of the "'white' socially dominant racialized group" (Thompson-Miller, 2017, page 21). Thus white individuals are not treated as pariah, even the poorest white Americans in Rural American have not faced discrimination on the basis of color in structural practices that are presumed as normal since the founding of the Americas, such as slavery. Racism is highly adaptable and when one has the ability to address underlying patterns linked to systemic racism, one can understand that "salvery didn't end; it evolved" as the human rights champion Bryan Stevenson observed (Alexander, 2010, xxxviii).

Therefore, "there is no black racism because there is no centuries-old system of racialized subordination and discrimiation designed by African Americans that excludes white Americans from full participation in the rights, privileges, and benefits of this society" (Thompson-Miller, 2017, page 21). According to Marcus Bell, however, the historical contradiction between US racial discourse and US racial practice has led to the systemic normalization of white superiority", thus the demand for systemic justice has been arising in movements in response (Thomson-Miller, 2017).

**Applied: Artificial intelligence in systemic injustices**

Artificial Intelligence, otherwise known as AI, denotes machines that can think autonomously, specifically software systems that are able to make decisions that normally require human level of expertise using key features of intentionality, intelligence, and adaptability. AI depends on algorithms to make decisions, that depend on machine learning. Machine learning is advanced data science where software can learn the input and analysis of passive or real-time data, find underlying trends, and come up with an outcome or decision over time. However, the excellence of AI is scored by the quality of adaptability, the ability to compile information and make the right decisions consistently and over the course of half a decade, the technology has been integrated into organizations and online patterns to make mild to critical decisions for humans such as considered who is considered for a job, who is more likely to succeed in university, who should go to jail and for how long, among other decisions (West, 2018).

"Since 1980 systemic work has expanded considerably in various directions (e.g. in artificial intelligence, child language development, discourse analysis and stylistics)...the very large systemic grammar now exists in computational form" (Koerner, 1996, page 276). "Dozen language-centered racism denial practices that sustain linguistic racial accommodation...*conceptual: colonization, conflation, extenuation, idealism, inflation, minimization, misdirection, non-definition, obfuscation, realism, rejection, and underdevelopment*" (Thompson-Miller, 2017, page 23). The use of such terms must be condemned through linguistic racial confrontation in ensuring a racial conceptual realism ideology does go unnoticed within the digital scope and development of machine learning algorithms used in artificial intelligence.

"Outputs from algorithms can and do reinforce biases and lead to disparate results because often times the datasets they rely on reflect historical biases and existing discrimination." (Bannan, 2020). Two of the most common tools which leverage AI are risk assessment algorithms and facial recognition. The former is often used in pre-trial stages during court trial on a defendant's chance of recidivism or failure to appear before a judge, probation and parole decisions, determining early release, etc. Facial recognition is often used by the point-of-entry of the criminal justice system by the police. Although face recognition is believed to have high classification accuracy of over 90%, landmark projects such as the 2018 "Gender Shades" and others have found that algorithms belonging to IBM, Microsoft, and Amazon "performed the worst on darker-skinned females and error rates of 34% higher than for lighter-skinned mails" (Najibi, 2020). For example, in 2018 Amazon built and tested an AI recruiting tool to help find and review potential job applicants and over time found the tool preferred lighter male-candidates over all female candidates for technical positions. Because the tool was built to identify patterns from previous resumes that made it into Amazon- which predominantly came from men- Amazon decided to renounce the tool. This shows the likeliness of unintended and faulty outcomes from AI that are not trained well enough and the need for risk-testing and auditing these algorithms before and after deployment (Bannan, 2020, page 12).

"Based on computer algorithms, AI machine learning are now the judge of who is innocent or who is held in jail. To put it simply, many would think isn't it better for a computer algorithm to decide your fate than a racist judge?" (Alexander, 2010, xxxvi). Such algorithms are prone to be biased due to its creator. And unless institutions systemically ensure the creators of these algorithms are following a modeled checklist set-up by policy to follow, and other strict regulations, only then can we prepare and systemically break apart this system before it goes far beyond our current world to the extent that mass incarceration defeats the purpose of what such future technologies and automations are capable of

Numerous studies conclude and suggest that institutions use dynamic face matcher selection, where face recognition systems are trained on different demographic cohorts, and is critical to reduce vulnerabilities on the performance of differentiating race or ethnicity cohorts (Brendan, 2012). Consequently, too address the issue, three senators introduced and passed the Algorithmic Accountability Act of 2019, which "requires the companies to test and fix flawed algorithms that result in biased or discriminatory outcomes"(Bannan, 2020, page 5) as well as those that (1) alter legal rights of consumers, (2) significantly impact consumers, and/or (3) involved the personal information of consumers on the basis of race, color, national origin, political opinion, religion, trade union membership, genetic data, biometric data, health, gender, gender identity, sexuality, sexual orientation, criminal convictions, or arrests ([Algorithmic

Accountability Act, 2019, page 6). However, such class action policy has yet to meet and regulate more specific industries that are using AI and machine learning algorithms, such as the criminal justice system, education system, and healthcare sector. A multi-stratal approach is required in order to prevent further exacerbation of racially discriminatory and privacy-invasive outcomes in systemic environments already suffering from systemic injustice.

## Chapter Two: Historical Systemic Patterns

The criminal justice and prison system is an amorphous system of more than 4,100 corporations, and their government conspirators profiting from mass incarceration of minority groups, especially that of black communities and individuals. The private sector, especially prisons and private housing distribution, had metastasized the carceral state. This method of systemic injustice began after the civil rights movement in the 1960s.

The Reagan-era privatization rampage was to essentially accomplish the reduction of the federal government and sector to the most extreme extent. It explicitly "embraced "privatization" as a new, supplementary strategy for minimalist government and deficit reduction" (Tingle, 1988). President Reagan's successors continued the "tough-on-crime" policies at harsher rates. In 1994 the Omnibus Crime Bill totaled 9 billion US dollars in prison expansion, and the Clinton Administration added more to the policy with the "truth-in sentencing", laws enacted to reduce the possibility of early release for incarceration and require offenders to stay a portion of the sentence before even becoming eligible for release. Consequently, between 1984-2005, a new prison or jail was built every 8.5 days in the U.S. with 70% of those built in rural communities suffering economic inequality.

Those who benefit are not just prison companies and government bodies on either end of the contract, but also the architects, engineers, and construction firms are profiting millions from laying a framework of mass incarceration by building the next iteration of the prison system. Corporations such as HDR, HOK, Turner Construction Co., Gilbane Building Co., Hensef Phelps, and McCarthy Building Companies are some of the largest firms working on expansion allowed by government contracts.

It wasn't until 2016 when the Obama administration announced a "phase out" of prison contracts after a study revealed they were less safe than public prisons. However, soon after the Trump administration took over, the "phase out" policy was rescinded in just weeks. Multiple of the private companies funded millions into his campaign and it shows "how private prison companies have faced significant volatility at the hand of political swings". Although the contracts seem to only deal with government officials and company executives, the discriminatory and inhumane practices go often unreported directly to the Department of Justice. From 1994 to 2017, only 69 formal investigations have been opened by the Civil Rights Special Litigation section that looks into pattern-and-practice of conduct of policing and discriminatory actions by private law enforcement in the U.S. (The Civil Rights Division's Pattern and Practice Police Reform Work, 2017, page 8).

**Privacy issues**

As of 2021, the patterns of expanding, privatizing, and profiting are now moving from physical spaces into digital space. Mass incarceration and time spent in prison are now enabled to follow former inmates through mass surveillance. GPS tracking devices are often shackled to released and returning citizens that are "provided by a private company that may charge up to $300 per month" and have movement restrictions in "open-air, digital prison" (Alexander, 2010).

## Chapter Three: Implicit Bias in Natural Language Processing

**Implicit bias & the race-blind justice system code**

Police and sheriff deputies are the point of entry into the criminal justice system and often the first of the institution's players that the public encounters. Further, the public encounter judges, defense lawyers, and prosecutors, all of which are able to sustain a system of criminality on the basis of racial morality and implicit bias. This is done primarily by labeling certain defendants as ""mopes", a construct that implies immorality" once a defendant is in the courtroom. Such a term refers to someone of lower socio-economic status and is lazy. Those who are labeled under these terms are typically charged with minor to non-violent crimes such as "possession of drugs and shoplifting, and "imply social dysfunction rather than criminal risk (Richardson, 2017, p. 870).

"Colorblind racism is more than just a 'doing' of rhetoric; it is a type of complicated habits that informs institutional practices and cultural memberships, and even aids in the organizational efficiency of the criminal courts...[This is] how professionals...'do racism' while 'doing justice" (Richardson, 2017, p. 872). Implicit bias is the unconscious impression towards a scenario or issue at hand. Because implicit bias is based on intuition, it can evolve into explicit and overt practice by the individual experiencing these thoughts because of systemic triage. This can be significantly problematic because it exacerbates racial inequalities to an extent that is inherently out of our reach to be personally hypervigilant, intent, and controlling of these mental heuristics. Unconscious associations that cause these heuristics "influence our perceptions, judgements, and behaviors without our conscious intent". This can include stereotyping and the attitudes we tie towards different racial groups. Such heuristics and bias seem too innate to be inevitable, this is especially true within environments where decision makers are limited on time and information. "Implicit racial bias can influence the discretionary decisions, perception, and practices of even the most well-meaning individuals in ways that are not readily observable" (Richardson, 2017, p. 876)

Implicit white favoritism is the theory that "the automatic association of positive stereotypes and attitudes with members of a favored group, leading to preferential treatment for persons of that group". This theory serves as a prime example of how implicit bias serves as a cue and trigger on how to respond to a certain stimulus in an environment, in this case it would be reasoning likeliness of crime on the basis of race. (Richardson, 2017, p. 876). According to Greenwald et al (2015), individuals of all races have implicit biases linking blacks with criminality over whites.

**Systemic Triage**

Our criminal justice system is modeled on choosing among the clients or defendants that are deserving of advocacy. This is apparent due to the negative correlation of insufficient amount of time and the excruciating number of nonviolent offenses that courtroom professionals are unequipped to handle.

> "[T]he provision of indigent defense is often likened to medical triage. Similar to hospital emergency rooms, [public defender] offices face demands that far outpace their resources. In order to save time to defend the cases that they find deserving, attorneys may plead out other cases quickly or go to trial unprepared. This reality means that for most [public defenders], the question is not "how do I engage in zealous and effective advocacy," but rather, "given that all my clients deserve aggressive advocacy, how do I choose among them?" (Richardson, 2017, 878).

The modern systemic triage model in our criminal justice system has come to its current state is largely due to policing practices that were enabled from the War on Drugs, which in turn increasingly allowed police to arrest an overwhelming number of people for irrelevant, minor, or non-violent crimes. During the War on Drugs, provisions were set out to alarmingly decrease restrictions on consent searches and seizures for law enforcement.

**Natural Language Processing (NLP) bias**

Natural Language Processing (NLP) is one of the two main machine learning research models that are used in AI algorithms. NLP is specifically a linguistic model and has a human-level achievement in language-understanding tasks or decision making. However, because this technology does not meet full human performance in complex thought, it does not know how to identify or reduce implicit bias. In terms of race bias, some machine learning models have outputted multiclass bias such as "black is to criminal as caucasian is to police'". Training data used by artificial intelligence algorithms include historically rooted data that show patterns of implicit racial bias, thus the algorithm is trained to replicate such when executing decision-making processes.

**Litigative Implications**

"[Artificial Intelligence] systems rely on copyrighted work for their factual nature and training data" (Levendowski, 2018). Much of the data is sourced from historically recorded work from secondary ownership, open-source databases, or government enclosed contracts that allow institutions or intergovernmental bodies to access information. However, algorithms focused on data that is retrieved from secondary ownership take more information than is used and do not follow a set of policy that keeps institutions accountable for the algorithm if discriminatory or racially biased outcomes occur. Besides copyright law, negligence law protects the privacy of individuals whose sensitive information is used. If utilized appropriately, negligence law can and should be used to increase internal surveillance of government systems and private entities leveraging such algorithms that use copyrighted or sensitive data.

If internal policy in the criminal justice system required courtrooms to take account quantitative data on how frequent, representative, and pervasive racialized practices are it could be used as a first step towards identifying discrimination. Internal accountability should be regularly rehearsed, especially within data, to inspect if there is a positive correlation between frequency of racialized practices based on reports and the outcomes of algorithms that are used in judicial decision-making processes. If and when positive correlations are determined and analysis proves that algorithms have discriminatory training data, policies should put in place requirements of rescinding the tool and data.

In recalling the tool and data, the next best practice for institutions is to follow the principles of agile delivery. An example of the principles is one set out by the UK Government Digital Service (GDS).

1. Discovery phase: Understanding the problem that needs to be solved.
2. Alpha phase: Building and testing different prototypes in response to problems outlined in the discovery phase.
3. Beta phase: Taking the best idea from the alpha phase and building a real tool for users.
4. Live phase: Running the new service in a sustainable way, and continuing to make improvements. (Eleanor, 2019)

**Chapter Four: Risk Assessment Algorithms and Policy**

The First Step Act (2018) requires the U.S. Attorney General to develop a "risk and needs assessment system" for the Federal Bureau of Prisons to assess each prisoner's risk of recidivism and determine what type of recidivism reduction programming appropriate for them" (Bannan, 2020, page 7). PATTERN (Prisoner Assessment Tool Targeting Estimated Risk and Need) is a system passed under Attorney General Willam P. Barr, alongside the Department of Justice, in response to the First Step Act. Although PATTERN is not classified as an artificial intelligence, it is an automated data informatics tool utilized by law enforcement that depends on biased historical data of discriminatory policing practices. For example, earlier in 2020, Attorney Barr ordered the release of a certain amount of federal prisoners to avoid superspreading the coronavirus. To determine which prisoner would get early release, the individual would need to score a minimum on PATTERN. A study found that white prisoners would be more favored over black prisoners because historical data, including that from the war on drugs era, showed black prisoners had more prison time than their counterparts. Artificial Intelligence systems are not equipped to quickly adapt to unprecedented times and neither is the technology advanced enough to identify what is or isn't systemically racist data.

Another prime example would be Northpointe. "Northpointe, the developer of the Correctional Offender Management Profiling for Alternative Sanctions (COMPAS), asserted that the system is not racially biased because it predicts overall recidivism equally well" for both race groups, white and black defendants. Most algorithms like COMPAS and PATTERN have two serious hurdles that allow for racial discrimination to appear in outcomes, one of which is the training data that is used to allow the AI system to adapt and the other as privacy and security risk (Bannan, 2020, page 13).

      Concentrating on the first hurdle, algorithms become what they are trained on or 'fed'. If the algorithm is fed training data collected during a time when discriminatory policing in Black communities was at its peak, the algorithm will consistently make the same policing decisions. This allows the systemic discrimination to evolve on into the digital scope, despite the new technology integrated. The case for COMPAS is that the data retrieved was from the 1980's to present patterns in policing. Due to the war on drugs and "tough-on-crime" policies initiated by Richard Nixon and several of his successors in the late 20th century, black communities were targeted prominently by CIA and FBI on cocaine crackdown and minor crime (Alexander, 2010). As laws have relaxed over the past 20 years on "tough-on-crime" policy, the previous data stays the same.

      Using an AI risk assessment on another automated AI corrective system has its own risks because conclusively the technology is not advanced enough to consume and train off of qualitative civic statutes instead of historically discriminatory data. "If risk assessments are used at all, they should only identify groups of people to be released immediately because automated recommendations for detention assume the guilt of the defendant" (Bannan, 2020, page 15). Risk assessments need more transparency in terms of the policies regulating them. Especially pre-trial assessments, which "must be transparent, independently validated, and open to challenge by an accused person's counsel". High risk systems need more mandated regulation as the data deals with more access to private and sensitive metrics such as race, gender, biometrics, political party, and criminal arrests. According to Bannan (2020), practices such as "data minimization, the retention period for personal information, and the ability of consumers to access and object to or correct the results of the" algorithm should be reinforced and passed by Congress to incorporate new technology and decision making strategies.

      Conclusively, companies need to reframe (1) what the clear point of the algorithms are

and (2) why such outputs were shown, as well as make public statements on the regulatory change taken. Civil rights groups are also suggesting that policies should be set forth to restrict algorithms' outcomes from becoming decisions and replaizing judges and counsels all together, and instead the outcomes could lead human decision makers to reach a final conclusion until algorithms are proven to have advanced decision ingenuity as humans.

## Conclusion

It has been speculated countless times throughout this dissertation that both public understanding of systemic discourse and full-implication of emerging technology in civic and litigative concerns has gained proper traction recently from the events of 2020. As a new form of system that allows law and institutions to use and perpetuate pre-existing racial disparities further into the digital world, it is suggested that raising awareness of the corrective terms that serve best when speaking about racial bias and inequity, especially systemic injustices, should be increasingly required within the law enforcement groups, courtrooms, corporation lawyers and algorithm engineers.

AI & RACIAL EQUITY                                                                                           24